\documentclass[sigconf,screen]{acmart}

\usepackage{booktabs} 
\usepackage[flushleft]{threeparttable}
\usepackage{xcolor}
\usepackage[utf8]{inputenc}
\usepackage{float}

\usepackage{amssymb}
\usepackage{amsmath}
\usepackage{ifthen}
\usepackage{caption}
\usepackage{hyperref}
\usepackage{array,graphicx}
\usepackage{soul}
\usepackage{balance}
\usepackage{pifont}
\usepackage{nicefrac}
\usepackage{mathtools}
\usepackage{tabularx}
\usepackage{xcolor,pifont}
\usepackage{tikz}
\usepackage{svg}
\usepackage{pdfpages}
\usepackage{array}
\usepackage{subcaption}
\usepackage{microtype}
\usepackage{sparklines}
\usepackage{booktabs}
\usepackage{graphicx}
\usepackage{longtable}
\usepackage{array}
\usepackage{xcolor}
\usepackage{supertabular}
\usepackage{booktabs}
\usepackage{xcolor}
\usepackage{adjustbox}
\usepackage{enumitem}

\newcommand{\rqbox}[1]{%
  \par\noindent
  \begingroup
    \setlength{\fboxrule}{0pt}
    \setlength{\fboxsep}{2pt}
    \colorbox{gray!10}{%
      \parbox{\dimexpr\linewidth-2\fboxsep\relax}{%
        \vspace{0.5pt}
        #1%
      }%
    }%
  \endgroup
  \par
}

\newcommand*\colourcheck[1]{%
  \expandafter\newcommand\csname #1check\endcsname{\textcolor{#1}{\ding{52}}}%
}
\newcommand*\colourcross[1]{%
  \expandafter\newcommand\csname #1cross\endcsname{\textcolor{#1}{\ding{56}}}%
}
\usepackage[vlined, boxruled, linesnumbered] {algorithm2e}
\SetKw{KwBy}{by}
\SetKw{KwBreak}{break}
\SetKw{KwReturn}{return}
\def\HiLi{\leavevmode\rlap{\hbox to \hsize{\color{gray!35}\leaders\hrule height .8\baselineskip depth .5ex\hfill}}}

\newlength\BARWIDTH
\newlength\BARHEIGHT
\newlength\BARWIDTHfour
\SetKwComment{Comment}{$\triangleright$\ }{}
\SetCommentSty{itshape}
\newboolean{showcomments}
\setboolean{showcomments}{true}
\ifthenelse{\boolean{showcomments}}
{\newcommand{\nb}[2] {
  \fcolorbox{black}{gray!20}{\bfseries\sffamily\scriptsize#1:}
  {\sf\small$\blacktriangleright$\textit{#2}$\blacktriangleleft$}
}
}
{\newcommand{\nb}[2]{}
}

\newcommand{\head}[1]{\noindent\textbf{#1.}}

\newcounter{fcounter}
\newcommand{\curl}[1]{\footnote{\url{#1}}}

\newcommand*\ChartFive[5]{%
  \begin{tikzpicture}[baseline={(current bounding box.south) - 10ex}]
    \foreach \i [count=\xi] in {#1, #2, #3, #4, #5} {%
      \draw[fill=blue, draw=white, line width=0.1mm]
        (\xi*\BARWIDTH-\BARWIDTH, 0) rectangle (\xi*\BARWIDTH, \i*\BARHEIGHT);
    }
    \draw[black] (0,0) -- (5*\BARWIDTH,0);
    \draw[black] (0,1.3*\BARHEIGHT) -- (5*\BARWIDTH,1.3*\BARHEIGHT);
  \end{tikzpicture}%
}

\newcommand*\ChartThree[3]{%
  \begin{tikzpicture}[baseline={(current bounding box.south) - 10ex}]
    \foreach \i [count=\xi] in {#1, #2, #3} {%
      \draw[fill=blue, draw=white, line width=0.1mm]
        (\xi*\BARWIDTH-\BARWIDTH, 0) rectangle (\xi*\BARWIDTH, \i*\BARHEIGHT);
    }
    \draw[black] (0,0) -- (3*\BARWIDTH,0);
    \draw[black] (0,1.3*\BARHEIGHT) -- (3*\BARWIDTH,1.3*\BARHEIGHT);
  \end{tikzpicture}%
}

\usepackage{tikz}

\expandafter\def\csname rqthreecolor@success\endcsname{green!70!black}
\expandafter\def\csname rqthreecolor@skip\endcsname{blue!70!black}
\expandafter\def\csname rqthreecolor@fail\endcsname{red!80!black}

\newcommand{\RQThreeColor}[1]{\csname rqthreecolor@#1\endcsname}

\makeatletter
\newcommand{\thickhline}{%
    \noalign {\ifnum 0=`}\fi \hrule height 1pt
    \futurelet \reserved@a \@xhline
}
\makeatother

\setcopyright{acmlicensed}
\copyrightyear{2026}
\acmYear{2026}
\acmDOI{XXXXXXX.XXXXXXX}
\acmConference[ASE '26]{Proceedings of the 41st IEEE/ACM International Conference on Automated Software Engineering}{October 12--16, 2026}{Munich, Germany}

\acmISBN{978-1-4503-XXXX-X/2026/06}

\title{Real-World Perturbation Testing of Autonomous Driving Systems}

\author{Stefano Carlo Lambertenghi}
\email{stefanocarlo.lambertenghi@tum.de}
\affiliation{%
  \institution{Technical University of Munich}
  \institution{fortiss GmbH}
  \country{Germany}
}

\author{Matthias Weil}
\email{matthias.weil@tum.de}
\affiliation{%
  \institution{Technical University of Munich}
  \country{Germany}
}

\author{Andrea Stocco}
\email{andrea.stocco@tum.de}
\affiliation{%
  \institution{Technical University of Munich}
  \institution{fortiss GmbH}
  \country{Germany}
}

\begin{document}

\begin{abstract}
Autonomous Driving Systems (ADS) must operate reliably under diverse conditions, yet representative data for rare or adverse scenarios is difficult to obtain. Perturbation-based testing is widely used to assess robustness, but most studies focus on offline datasets or simulation, leaving open questions about how such results translate to real-world driving.
We present a large-scale study of 72 camera and LiDAR perturbations, evaluated across three testing modalities: offline model-level analysis, hardware-in-the-loop execution, and closed-loop system-level testing on a full-scale autonomous vehicle. The study covers both an end-to-end vision-based driving model and a modular LiDAR-based perception and planning stack.

Our results reveal a clear gap between testing levels. For camera-based systems, perturbations with limited offline impact can still induce unstable control and failures in real-world driving. For LiDAR-based systems, degradation is more consistent at the perception level but weakly predictive of system-level failures. Across both modalities, model-level metrics alone are insufficient to identify the most harmful perturbations.
We further show that real-time feasibility is a key constraint in real-world testing, and that robustness observations obtained from recorded data do not consistently transfer to closed-loop behavior on a physical vehicle, highlighting the importance of complementary real-world, system-level evaluation.
\end{abstract}

\maketitle
\section{Introduction}\label{sec:introduction}

Autonomous Driving Systems (ADS) rely on perception pipelines built on heterogeneous sensors such as cameras and LiDAR to interpret dynamic environments and support downstream driving decisions~\cite{survey-lei-ma,yurtsever2020survey,grigorescu2020survey,li2024panopticperceptionautonomousdriving}. Ensuring robustness across diverse operating conditions remains difficult, since even small input perturbations can degrade perception and propagate to unsafe system-level behaviour, especially under out-of-distribution (OOD) conditions~\cite{DBLP:journals/corr/abs-1912-12162,hendrycks2019benchmarkingneuralnetworkrobustness,dodge2016understanding,geirhos2020generalisation}.

One of the main limitations is the lack of representative data. Rare but safety-critical conditions, such as adverse weather, illumination changes, sensor noise, and environmental artifacts, are difficult to collect systematically and are often underrepresented in training and evaluation datasets~\cite{7823109,sharma2024survey}, limiting testing effectiveness. 

To address this problem, prior work has widely explored \emph{perturbation testing}, where controlled transformations are applied to sensor inputs to emulate challenging conditions~\cite{hendrycks2019benchmarking,hendrycks2020augmix,rusak2020simple,Laermann-2019,8388338}. In computer vision, benchmarks such as ImageNet-C, ImageNet-P, and MNIST-C established perturbation-based robustness evaluation as a standard offline practice~\cite{hendrycks2019benchmarking,hendrycks2019benchmarkingneuralnetworkrobustness,mu2019mnistc}, while augmentation methods such as RandAugment and AugMix improve robustness through adversarial training~\cite{cubuk2019randaugment,hendrycks2020augmix}. In autonomous driving, perturbations have been applied to camera images~\cite{deepxplore,deeptest,deeptest,deepbillboard} and LiDAR point clouds~\cite{dong2023benchmarking,dong2023benchmarkingrobustness3dobject,10588664}, but these studies use static datasets, where downstream driving effects cannot be observed~\cite{survey-lei-ma,briand-offline-emse,2023-Stocco-EMSE}.

Recent work has integrated perturbations into simulation-based testing to enable closed-loop evaluation~\cite{NeelofarTOSEM,NeelofarICSE,pafot,epitester,lu2023deepqtesttestingautonomousdriving,2025-Lambertenghi-ICST,2023-Stocco-TSE,dataAugment2020Liu,yoon2023learning,DeepManeuver,ayerdi2023metamorphic,Luan2023Efficient}. 
These approaches better capture temporal effects but remain constrained by simulation fidelity~\cite{zhao2025statistical,2024-Lambertenghi-ICST,2025-Lambertenghi-ICST}. It therefore remains unclear whether perturbation-induced degradations observed offline translate into failures during real-world closed-loop driving~\cite{2020-Haq-ICST,briand-offline-emse,2023-Stocco-EMSE}, despite continued evidence of incidents highlighting the importance of real-world testing~\cite{damon2018uber,neal2017tesla,david2018uber,stempel2025tesla,jamali2025tesla,reuters2026waymo}.

To address this, we present a comprehensive evaluation of camera and LiDAR perturbations, benchmarking a wide range of techniques from the literature across two ADAS tasks of increasing complexity, a vision-based end-to-end driving model, and a modular LiDAR-based perception and planning stack, and three testing modalities: offline model-level evaluation on static datasets, hardware-in-the-loop (HiL) testing on the vehicle compute stack, and vehicle-in-the-loop (ViL) testing on a real autonomous vehicle.

First, we perform model-level evaluation on recorded real-world datasets to quantify how perturbations affect model outputs and task-specific metrics. Second, we deploy perturbations in a HiL setup to assess real-time feasibility on automotive hardware. Third, we inject perturbations into live sensor streams during closed-loop driving on a full-scale autonomous vehicle to directly observe their impact on real driving behaviour. We also evaluate whether fine-tuning with perturbed data improves the robustness of the end-to-end driving model under real-world OOD conditions.

Our results show clear differences across modalities and testing levels. Camera perturbations often induce high-variance control deviations: low average offline errors can still lead to unstable driving. LiDAR perturbations instead produce more consistent degradation, mainly reducing detection retention while preserving comparatively stable localization. We also find limited transfer across testing levels: perturbations that appear mild offline can still trigger lane departures or collisions in closed-loop driving, while others that strongly affect model-level metrics have limited observable system-level impact. Most perturbations satisfy real-time constraints on the tested automotive hardware, and fine-tuning with perturbed data improves robustness under naturalistic perturbations such as rain, glare, and sensor artifacts while preserving nominal performance.

Our paper makes the following contributions:

\begin{description}[noitemsep,nolistsep]

\item [Framework] A ROS-based framework of 72 perturbations for conducting robustness testing across model-level, HiL, and ViL deployment for camera and LiDAR sensors.

\item [Empirical Study] To the best of our knowledge, this is the first software engineering paper to conduct cross-modality, cross-level robustness analysis in the context of real-world testing of ADS at the system-level with a real-world full-size vehicle, on end-to-end and modular driving models.

\item [Findings] We identify a gap between model-level robustness and system-level behavior, showing that perturbations with limited offline impact can still induce failures in closed-loop driving, and that real-time feasibility is a critical constraint for real-world system-level ADS testing.

\end{description}
\section{Image and LiDAR Perturbations}\label{sec:pert}

\begin{figure}[t]
    \centering
    \includegraphics[width=1\linewidth]{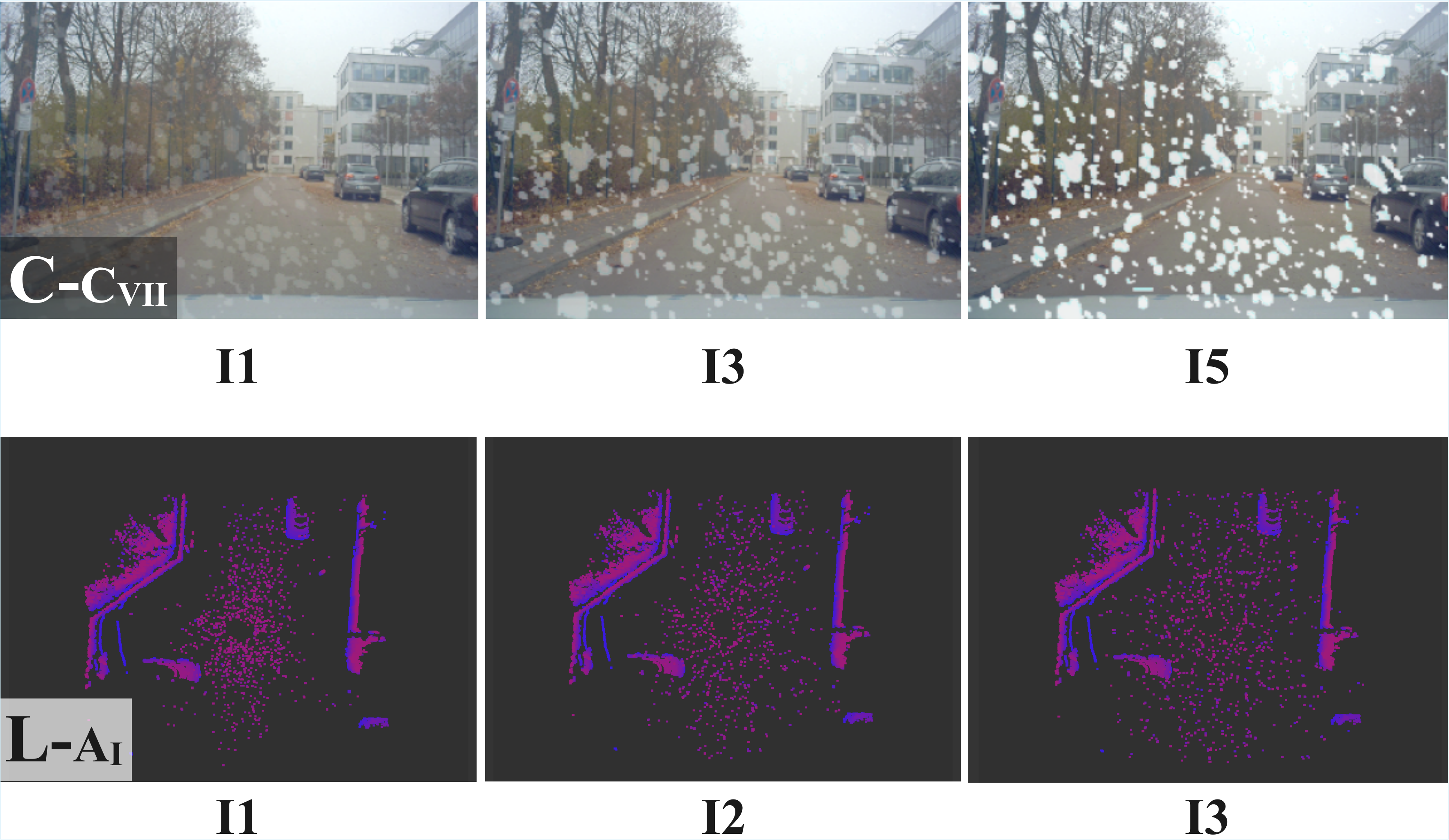}
    \caption{Examples of camera (top) and LiDAR (bottom) perturbations illustrating the Snow effect at different intensities.}
    \label{fig:perts}
\end{figure}

We study 72 perturbations from the literature for both camera images and LiDAR point clouds, the two most important ADS sensors~\cite{survey-lei-ma}.
For simplicity of exposure, in this paper, we assign a compact identifier to each perturbation: camera perturbations are denoted as \textbf{C-X\textsubscript{Y}}, whereas LiDAR perturbations are denoted as \textbf{L-X\textsubscript{Y}}, where \textbf{X} identifies the perturbation category and \textbf{Y} an ordinal index within that category. Examples of camera and LiDAR perturbations are shown in \autoref{fig:perts}.

\subsection{Camera Perturbations}

For camera inputs, we adopt the perturbation suite from PerturbationDrive~\cite{2025-Lambertenghi-ICST,2026-Leonhard-SCP}, including 41 image corruptions that emulate environmental effects, sensor artifacts, and visual distortions. Each perturbation is applied with five severity levels following the configuration of the original paper. More in detail, we evaluate the following perturbations.

\head{Noise Perturbations (A)}
Noise perturbations simulate sensor noise and signal interference: Gaussian (C-A\textsubscript{I}), Poisson (C-A\textsubscript{II}), impulse (C-A\textsubscript{III}), JPEG compression (C-A\textsubscript{IV}), and speckle noise (C-A\textsubscript{V})~\cite{dodge2016understanding,geirhos2020generalisation,hendrycks2019benchmarking,rusak2020simple,Laermann-2019,ayerdi2023metamorphic,Luan2023Efficient,michaelis2020benchmarking,mu2019mnistc}.

\head{Blur Perturbations (B)}
Blur perturbations model optical and motion artifacts: defocus (C-B\textsubscript{I}), glass (C-B\textsubscript{II}), motion (C-B\textsubscript{III}), zoom (C-B\textsubscript{IV}), Gaussian (C-B\textsubscript{V}), low-pass (C-B\textsubscript{VI}), and high-pass filtering (C-B\textsubscript{VII})~\cite{hendrycks2019benchmarking,rusak2020simple,Luan2023Efficient,mu2019mnistc,michaelis2020benchmarking,geirhos2020generalisation}.

\head{Weather and Illumination Perturbations (C)}
These perturbations emulate environmental and illumination effects: frost (C-C\textsubscript{I}), snow (C-C\textsubscript{II}), fog (C-C\textsubscript{III}), brightness (Increase: C-C\textsubscript{IV}, Decrease: C-C\textsubscript{V}), contrast (C-C\textsubscript{VI}), rain (C-C\textsubscript{VII}), raindrops (C-C\textsubscript{VIII}), lightning (C-C\textsubscript{IX}), smoke (C-C\textsubscript{X}), sun glare (C-C\textsubscript{XI}), and snowflakes (C-C\textsubscript{XII})~\cite{hendrycks2019benchmarking,rusak2020simple,ayerdi2023metamorphic,Luan2023Efficient,michaelis2020benchmarking,cubuk2019randaugment}.

\head{Distortion Perturbations (D)}
Distortion perturbations modify pixel geometry: elastic deformation (C-D\textsubscript{I}), pixelation (C-D\textsubscript{II}), sample pairing (C-D\textsubscript{III}), and sharpening filters (C-D\textsubscript{IV})~\cite{hendrycks2020augmix,ayerdi2023metamorphic,cubuk2019randaugment}.

\head{Graphic Pattern Perturbations (E)}
Graphic pattern perturbations overlay synthetic structures: splatter (C-E\textsubscript{I}), dotted lines (C-E\textsubscript{II}), zigzag patterns (C-E\textsubscript{III}), Canny edges (C-E\textsubscript{IV}), cutout (C-E\textsubscript{V}), and object overlays (C-E\textsubscript{VI}).

\head{Color and Tone Adjustments (F)}
Color and tone perturbations modify chromatic properties: false color (C-F\textsubscript{I}), phase scrambling (C-F\textsubscript{II}), histogram equalization (C-F\textsubscript{III}), white balance (C-F\textsubscript{IV}), grayscale (C-F\textsubscript{V}), saturation (C-F\textsubscript{VI}), and posterization (C-F\textsubscript{VII})~\cite{geirhos2020generalisation,cubuk2019randaugment,hendrycks2020augmix}.

\subsection{LiDAR Perturbations}

For LiDAR inputs, we adopt a total of 31 perturbation models from 3D-Corruptions-AD~\cite{dong2023benchmarking} and MultiCorrupt~\cite{10588664}, which provide physically grounded and data-driven corruptions for point clouds. These methods cover weather effects, sensor artifacts, noise processes, and motion-related distortions. Each perturbation is applied with three severity levels following their original implementation.

In addition to these methods, we introduce lightweight approximations for selected perturbation types (i.e., weather effects and dynamic artifacts). These approximations are designed to capture the same qualitative degradation patterns as their simulation-based counterparts, while being computationally cheaper for real-world execution.

\head{Weather and Environment Perturbations (A)}
They are modeled using both simulation-based and approximate methods.
Snow includes two physically grounded snowfall models from 3D-Corruptions-AD (L-A\textsubscript{I} and L-A\textsubscript{II}) and a variant from MultiCorrupt (L-A\textsubscript{III}). These approaches are based on light scattering formulations for LiDAR sensing~\cite{HahnerCVPR22}. In addition, we include a lightweight scattering approximation (L-A\textsubscript{IV}) that probabilistically attenuates and removes points based on distance and intensity.
Rain includes a simulation-based rainfall model from 3D-Corruptions-AD (L-A\textsubscript{V}) and a lightweight attenuation model (L-A\textsubscript{VI}) that reduces return intensity and sparsifies distant points.
Fog includes a physically based model from 3D-Corruptions-AD (L-A\textsubscript{VII}), a variant from MultiCorrupt (L-A\textsubscript{VIII}), and a lightweight attenuation approximation (L-A\textsubscript{IX}) based on exponential decay of signal strength with distance.
Strong sunlight (L-A\textsubscript{X}), adopted from 3D-Corruptions-AD, introduces background noise and spurious returns.

\head{Sensor and Field-of-View Artifacts (B)}
Density decrease includes global point thinning from 3D-Corruptions-AD (L-B\textsubscript{I}) and a stochastic reduction variant from MultiCorrupt (L-B\textsubscript{II}). Beam reduction (L-B\textsubscript{III}) simulates reduced LiDAR channels and is adopted from MultiCorrupt. Cutout (L-B\textsubscript{IV}) removes regions from the point cloud and is adopted from 3D-Corruptions-AD. Crosstalk noise (L-B\textsubscript{V}) and field-of-view filtering (L-B\textsubscript{VI}) are also from 3D-Corruptions-AD.

\head{Global Noise Perturbations (C)}
They include Gaussian noise (L-C\textsubscript{I}), uniform noise (L-C\textsubscript{II}), and impulse noise (L-C\textsubscript{III}), all from 3D-Corruptions-AD.

\head{Local Corruptions (D)}
They include density reduction (L-D\textsubscript{I}), cutout (L-D\textsubscript{II}), and localized Gaussian (L-D\textsubscript{III}), uniform (L-D\textsubscript{IV}), and impulse noise (L-D\textsubscript{V}), all from 3D-Corruptions-AD.

\head{Affine Transformations (E)}
They include shear (L-E\textsubscript{I}), scaling (L-E\textsubscript{II}), and rotation (L-E\textsubscript{III}), from 3D-Corruptions-AD.

\head{Motion and Alignment Perturbations (F)}
Motion blur (L-F\textsubscript{I}) introduces temporal smearing of points and is adopted from MultiCorrupt. Moving-object perturbations include dynamic noise injection from 3D-Corruptions-AD (L-F\textsubscript{II}) and ghost point generation implemented by us (L-F\textsubscript{III}). Spatial misalignment (L-F\textsubscript{IV}) applies rigid transformations and is from MultiCorrupt.

\section{Evaluation Framework}\label{sec:framework}

Our evaluation framework includes a real autonomous driving vehicle, a Volkswagen Passat Variant GTE plug-in hybrid~\cite{8813784},\footnote{Reference anonymized for double-blind review.} along with a software library that unifies camera and LiDAR perturbations from multiple sources (e.g., PerturbationDrive, 3D-Corruptions-AD, MultiCorrupt) under a unified execution interface, described next.

\subsection{Vehicle Platform}

\begin{figure}
    \centering
    \includegraphics[width=1\linewidth]{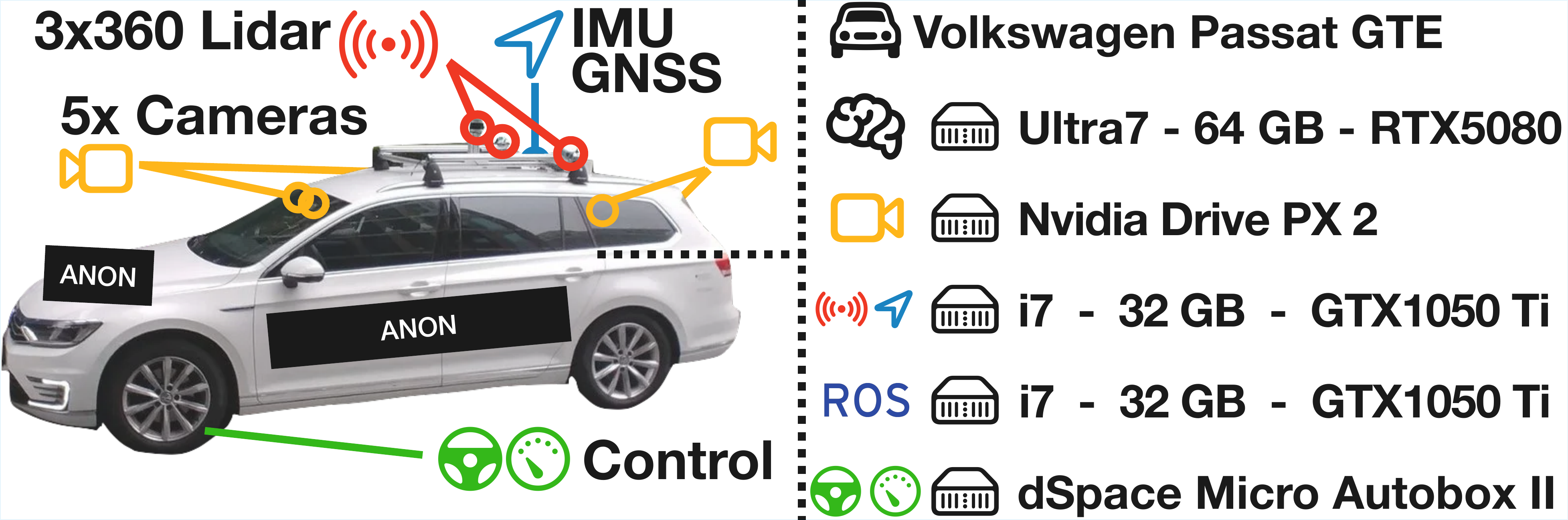}
    \caption[\textit{fortuna} vehicle]{Autonomous driving platform used in this work: sensors (left) and onboard computing hardware (right).}
    \label{fig:fortuna}
\end{figure}

As illustrated in \autoref{fig:fortuna} (left), the vehicle is equipped with a sensor suite providing a $360^{\circ}$ field of view. The setup includes five Sekonix automotive cameras (2.3 MP), consisting of two front-facing cameras ($60^{\circ}$ and $120^{\circ}$), two side-facing cameras ($120^{\circ}$), and one rear camera ($120^{\circ}$). The vehicle is also equipped with three Velodyne LiDAR sensors mounted on a roof rack, including a central 32-layer VLP-32C (range up to 200 m) and two tilted VLP-16 sensors covering the vehicle's sides. For localization, the platform uses an iMAR iNAT FSSG 1 GNSS INS system with RTK, providing positioning accuracy of approximately 2 cm, used both by modular driving pipelines and for recording high-precision telemetry during experiments.

The onboard computing infrastructure (\autoref{fig:fortuna} right) includes two CAR PCs (Intel Core i7, 32 GB RAM, NVIDIA GTX 1050 Ti), an NVIDIA Drive PX2 used for perception processing, and a high-performance computing unit dedicated to onboard intelligence and perturbation execution, equipped with an Intel Ultra 9 processor, an NVIDIA RTX 5080 GPU, and 64 GB of memory.
Low-level vehicle control is handled by a dSPACE MicroAutoBox~\cite{dSPACE_2026}, which interfaces with the vehicle CAN bus through a secure gateway to control steering and acceleration actuators in real time. All computing units are interconnected through a secure Ethernet local network.

\subsection{Testing Modalities}

\subsubsection{Model-Level}
Perturbations are applied offline to recorded sensor data, i.e., without ROS or vehicles integration. Camera perturbations modify image frames before inference, while LiDAR perturbations affect point clouds prior to perception processing, enabling batch execution and isolating input-level effects.

\subsubsection{Hardware-in-the-Loop}
Recorded ROS sensor messages are replayed on the vehicle hardware, and perturbations are applied online within the same ROS pipeline used in the live system. This enables evaluation of runtime performance under realistic conditions and filtering of perturbations that violate real-time constraints.

\subsubsection{Vehicle-in-the-Loop}

Perturbations are injected online while the systems under test run on the real vehicle. ROS nodes subscribe to raw sensor topics, apply perturbations in real time, and republish modified data to downstream modules. The same nodes used in HiL are reused on live sensor streams.
Perturbations are injected seamlessly at the sensor interface via Ethernet, requiring no modifications to the ADS.

\subsection{Perturbation Library}

We extended PerturbationDrive~\cite{2025-Lambertenghi-ICST} beyond its original image-based, simulation-driven setting by incorporating LiDAR perturbations and enabling execution across model-level, HiL, and real-world ViL environments. 

\head{Camera perturbations}
Camera perturbations operate on image tensors. In offline execution, they are applied before inference; in ROS-based settings, a node subscribes to image topics, applies transformations, and republishes frames while preserving message structure and timestamps.
For dynamic effects (e.g., rain, smoke), spatial masks and particle structures are pre-computed and reused to reduce runtime overhead.

\head{LiDAR perturbations}
In offline execution, they are applied directly to recorded data; in ROS-based settings, a node processes streaming point clouds, converts them to internal representations, applies perturbations, and republishes the results.
To handle high data rates, the node adopts a decoupled producer--consumer design: incoming messages are buffered via non-blocking callbacks, while perturbations are applied in a separate processing loop using vectorized operations, preventing latency propagation.
In particular, physics-based perturbations from MultiCorrupt are adapted to the Velodyne VLP-32C~\cite{VelodyneWiki} by reconfiguring sensor-specific parameters (e.g., beam count, vertical field of view, angular resolution, range limits) and regenerating scattering lookup tables.

Preliminary experiments showed that full physics-based models incur prohibitive overhead for real-time execution. We therefore adopt phenomenological approximations of weather effects that preserve key properties (e.g., distance-dependent attenuation and near-field backscatter) while reducing computation to constant time per point.
Atmospheric attenuation is modeled using a stochastic formulation derived from the Beer--Lambert law~\cite{swinehart1962beer}. 
For a point at distance $d$, the survival probability is: $P_{\text{surv}}(d) = e^{-\alpha d}$, where $\alpha$ controls perturbation severity. Points are stochastically discarded according to $P_{\text{surv}}$, simulating distance-dependent signal loss observed in fog, rain, and snow. Synthetic back-scatter noise is injected in the near-field region to emulate particle reflections, reproducing characteristic ghost returns. This formulation enables execution above 5Hz without per-ray simulation.

The LiDAR node supports both global and actor-specific perturbations. For actor-specific perturbations, the node first identifies points associated with dynamic objects using 3D bounding boxes generated in real time from the unperturbed stream, and then applies the corruption only to those points. This mode is used to enable \emph{Local Corruptions (D)} and moving-object perturbations in \emph{Motion and Alignment Perturbations (F)}.
As defined in the original libraries, camera perturbations use five severity levels, whereas LiDAR perturbations provide either three or five; for consistency, we unify all LiDAR perturbations to a three-level scale.
\section{Empirical Study}\label{sec:empirical-study}

\subsection{Research Questions}\label{sec:research_questions}

\noindent
\textbf{RQ\textsubscript{1} (model-level testing):}
\textit{How do camera and LiDAR perturbations affect the outputs of ADS during model-level testing?}

\noindent
\textbf{RQ\textsubscript{2} (hardware-in-the-loop feasibility):}
\textit{Can perturbation nodes meet the real-time constraints of the vehicle platform during hardware-in-the-loop testing?}

\noindent
\textbf{RQ\textsubscript{3} (system-level testing):}
\textit{How do perturbations affect driving behavior when injected into live sensor streams during ViL testing?}

\noindent
\textbf{RQ\textsubscript{4} (perturbation-based fine-tuning):}
\textit{Does training an end-to-end driving model on perturbed data improve robustness under real weather conditions?}

RQ\textsubscript{1} isolates the direct impact of input perturbations by applying them offline to recorded sensor data, enabling controlled, repeatable evaluation of model robustness independent of system-level dynamics.
RQ\textsubscript{2} evaluates whether perturbations that are effective offline remain feasible under real-time constraints on automotive hardware, a prerequisite for hardware-in-the-loop and on-vehicle testing.
RQ\textsubscript{3} assesses whether perturbation effects observed in controlled settings translate to real-world driving behavior. While model- and hardware-level evaluations capture functional and timing aspects, only closed-loop, on-vehicle execution reveals their impact on system-level performance under realistic dynamics and feedback.
RQ\textsubscript{4} evaluates whether perturbations can be used not only for testing but also to improve robustness. By fine-tuning on perturbed data and assessing performance under real weather conditions, this RQ examines the transferability of synthetic perturbations to real-world robustness gains.

\subsection{Driving Scenario}

\begin{figure}[t]
    \centering
    \includegraphics[width=1\linewidth]{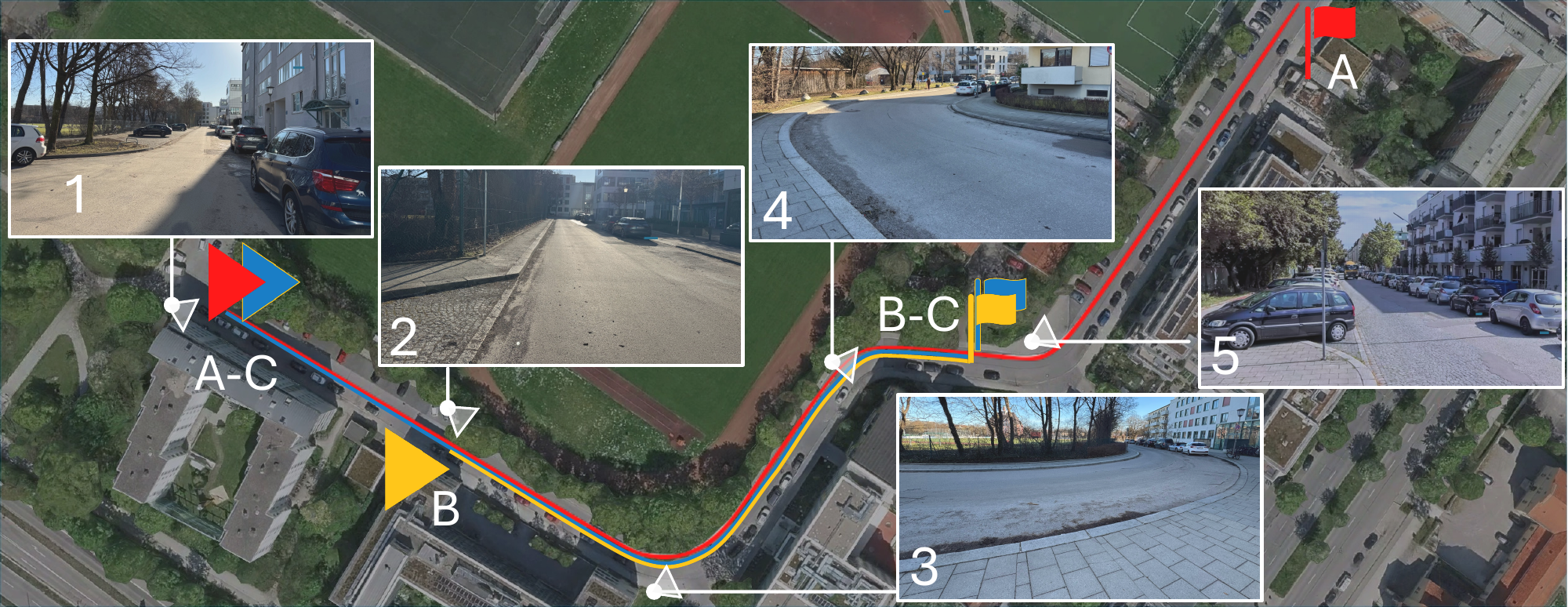}
    \caption{Real-world testing scenario.}
    \label{fig:driving_scenario}
\end{figure}

Experiments are conducted in \textbf{anonymized}, a public urban street (450 m, two-way, 30 km/h, \autoref{fig:driving_scenario}) with sidewalks, no lane markings, parked vehicles partially occupying the right driving lane, and three curves, but no intersections or pedestrian crossings. This yields a controlled yet non-trivial scenario, as the vehicle must infer the drivable corridor without lane markings while handling road curvature and lateral constraints induced by parked vehicles.
We define three routes: Route A (full route) is used for training, dataset collection, and fine-tuning tests (RQ\textsubscript{1}, RQ\textsubscript{4}). Route B starts 70 m after Route A and includes two turns; it is used for camera-based experiments (RQ\textsubscript{3}). Route C shares the start of Route A and ends at Route B's endpoint; it is used for LiDAR-based experiments (RQ\textsubscript{3}).

\subsection{Objects of Study}\label{sec:objects_of_study}

\subsubsection{End-to-end Driving Model}

The system under test is a vision-based driving model inspired by DAVE-2~\cite{bojarski2016end} that predicts continuous steering commands from RGB images and comprises five convolutional layers followed by three fully connected layers, with Batch Normalization and Dropout to improve training stability and generalization.
Input frames ($800 \times 503$) are cropped by 204 pixels from the top and 35 from the bottom, yielding $800 \times 264$ images focused on the road surface.
The training dataset includes 150 driving runs and 292k image-command pairs, and is augmented via horizontal flipping with inverted steering commands, following existing guidelines~\cite{bojarski2016end}. Data is split into 80\% training and 20\% validation with fixed seeds for reproducibility. 
The model is trained with Adam~\cite{kingma2017adammethodstochasticoptimization}, a learning rate of $1\mathrm{e}{-4}$, batch size 16, and MSE loss, for up to 100 epochs with early stopping (patience 10).

\subsubsection{Modular Perception Model}

The system uses the LiDAR-based Autoware Mini pipeline~\cite{autoware_mini}, a Python port of the Autoware stack~\cite{autoware-foundation-github}, deployed both on a Lexus RX450h at the University of Tartu~\cite{adlvehicle} and adapted to our vehicle.
Perception is based on geometric clustering: ground points are removed, and the point cloud is segmented using DBSCAN clustering~\cite{dbscan}, with clusters converted into 3D bounding boxes. The ADS includes perception, global and local planning, and control. Global planning uses a Lanelet2 map~\cite{lanelet2}, while local planning adapts trajectories online and supports street-level navigation around parked vehicles. Trajectories are tracked using a Pure Pursuit controller.

\subsection{Procedure and Metrics}

\subsubsection{RQ\textsubscript{1}}

We evaluate perturbation impact on the two ADS offline using a testing dataset. A separate recording on Route A was performed under nominal driving conditions. The recording consists of a 128-second cloudy-weather sequence with 3840 images and 640 point clouds. For each perturbation, perturbed versions of the dataset are generated at all intensity levels.

For camera inputs, perturbations are first applied before DAVE-2 inference, after which steering predictions are compared against the unperturbed baseline. We quantify their effect using Mean Squared Error (MSE), reporting the average and maximum across severity levels. MSE is a standard steering-angle regression metric and is also used in the original DAVE-2 implementation~\cite{bojarski2016end}. We additionally report the maximum steering deviation, i.e., the maximum absolute difference from the baseline prediction.

For LiDAR inputs, perturbations are first applied to point clouds before perception, after which detections are compared against the unperturbed baseline. We quantify the effect using Retention Rate (Recall) of baseline detections~\cite{kitti,lidar-metrics,Recall-retrate}, reporting the average and minimum across severity levels, and Average Translation Error (ATE)~\cite{caesar2020nuscenes,lidar-metrics}, reporting its maximum across severity levels. Recall is the percentage of baseline detections that remain detectable after perturbation, while ATE measures translation error between matched bounding boxes in perturbed and baseline point clouds.

\subsubsection{RQ\textsubscript{2}}

We integrate the perturbation library on the vehicle’s onboard hardware within the ROS perception pipeline. Each perturbation is benchmarked over a 10-second window, recording perturbation latency and ROS node overhead for each frame or point cloud.
We measure average latency (mean processing time), maximum latency (worst case across severity levels), and node overhead, i.e., the additional latency introduced by the ROS node. 

This separation lets us assess whether middleware contributes to real-time violations.
Latency is computed per severity level and aggregated per perturbation as the mean across severity levels and the maximum observed value.
For ViL evaluation in RQ\textsubscript{3}, camera perturbations must remain below 33,ms and LiDAR perturbations below 200,ms; others are excluded.

\subsubsection{RQ\textsubscript{3}}

Perturbations are evaluated starting from the highest severity level ($5$). Upon failure, lower severities ($[4..1]$) are tested. This prioritizes failure-inducing conditions and reduces real-world testing cost. All experiments are conducted at a fixed speed of 20 km/h.
All trajectory-based metrics are computed at the run level and aggregated per perturbation using only non-failing runs. In addition, for each perturbation, we summarize the run outcome separately for each tested intensity level. 

Each experiment is classified as \textit{success} or \textit{failure}, with results summarized over five severity levels for the camera-based system and three representative levels for the LiDAR-based system. For all runs, we report completion rate, i.e., the percentage of the scenario completed before failure~\cite{2025-lambertenghi-ASE}.
Failures are further categorized into five types. Erratic start (\textbf{E}) refers to unstable behavior immediately after initialization, characterized by large steering oscillations ($>0.3$ between consecutive frames) that prevent starting experiments. Understeer (\textbf{U}) denotes insufficient steering during turns, causing the vehicle to drift outward. Oversteer (\textbf{O}) refers to excessive steering leading to sharp deviations from the intended trajectory. Out of road (\textbf{R}) describes lane departure during straight segments due to accumulated lateral error. Collision (\textbf{C}) corresponds to contact with static obstacles (e.g., parked vehicles), typically due to perception or planning failures.

We also compute cross-track error (CTE)~\cite{2023-Stocco-TSE,cte}, defined as the per-frame lateral distance between the driven trajectory and a reference region constructed from three nominal trajectories. CTE is computed over the full trajectory for non-failing runs and up to the failure point for failing runs. In the latter case, the final $d$ meters ($d = 5\mathrm{m}$) before failure are excluded, and at least $d+1\mathrm{m}$ of valid trajectory must remain after truncation, in addition to a minimum completion threshold. This ensures that CTE reflects stable driving behavior rather than terminal instability immediately before failure.

For the end-to-end model, control stability is evaluated using steering jitter, i.e., the mean absolute change between consecutive steering commands. This metric has been used to measure driving stability in end-to-end systems~\cite{2025-lambertenghi-ASE,jitter}, including PerturbationDrive~\cite{PerturbationDrive}.
For the LiDAR-based ADS, we measure near-field detection coverage, i.e., the average proportion of vehicle detections maintained within 5,m during closed-loop operation. This distance slightly exceeds our vehicle length (4.7,m), covering the immediate obstacle-avoidance region.

\subsubsection{RQ\textsubscript{4}}

We fine-tune DAVE-2 on an augmented training set obtained by applying camera perturbations to the original images. For each sample, one perturbed variant is generated by randomly sampling a perturbation type and severity level from those used in RQ\textsubscript{1}--RQ\textsubscript{3}. The augmented data is combined with the original dataset, approximately doubling its size while retaining clean samples to preserve nominal performance. The model architecture, training procedure, and data splits remain unchanged to ensure a fair comparison.

The fine-tuned model is deployed on the vehicle and evaluated using the same closed-loop driving setup as in RQ\textsubscript{3}. In contrast to RQ\textsubscript{3}, which uses reduced routes, the evaluation in RQ\textsubscript{4} is conducted on the full driving scenario (Route~A) to assess robustness under more diverse and extended driving conditions.
First, both models are tested under nominal conditions to verify that the fine-tuned model maintains comparable performance in the absence of perturbations. Second, the models are evaluated under naturalistic environmental disturbances, including mud on the windshield, raindrops, sunlight glare, wet road surfaces, and heavy cloud coverage.

Each condition is evaluated in three runs on Route~A. The evaluation follows the same metrics as in RQ\textsubscript{3}, including run outcome, completion rate, and failure type. For successful runs, we report cross-track error and steering jitter as defined in RQ\textsubscript{3} to assess trajectory similarity and control stability.

\subsubsection{Summary}
Our experiments comprise more than 740 test executions: 298 at the model level, 298 in the HiL setup, 119 in the ViL setup, and 30 for fine-tuning. In total, the study covers more than a million ADS predictions and over 48 hours of real-world driving conducted across multiple months to capture diverse weather conditions, with collected ROS data exceeding 600\,GB. Across model-level, HiL, and ViL experiments, the total runtime exceeds 80 hours.

\input{table-perturbations-clustering-only}

\subsection{Results}

\subsubsection{Model-level testing (RQ\textsubscript{1})}

The first four columns of \autoref{tab:rq1-combined} report model-level robustness metrics for camera and LiDAR perturbations.
For the camera model, 32/42 perturbations show a clear monotonic increase in error with severity. The remaining cases follow a few recurring exception patterns: \textit{C-C\textsubscript{VI}, C-B\textsubscript{IV}}, and \textit{C-B\textsubscript{VII}} remain nearly flat across severity, \textit{C-E\textsubscript{IV}} shows an inverted trend, and \textit{C-E\textsubscript{III}} and \textit{C-C\textsubscript{IX}} peak at intermediate severities. The remaining four, such as \textit{C-F\textsubscript{I}}, are broadly increasing but locally irregular. 
In terms of magnitude, only 8/42 camera perturbations exceed 1.0 in Avg.\ MSE, whereas 14/42 exceed 1.0 in Max.\ MSE, showing that peak degradation is more common than sustained average degradation. The strongest cases are \textit{C-C\textsubscript{XII}, C-D\textsubscript{III}, C-F\textsubscript{I}, and C-E\textsubscript{V}}, all with high average error (Avg.\ MSE $>1.6$) and high peak error (Max.\ MSE $>3.0$). By contrast, most perturbations remain below 1.0 in Avg.\ MSE (34/42) and Max.\ MSE (28/42). A large gap between Avg.\ MSE and Max.\ MSE indicates degradation concentrated at one or a few higher severity levels rather than across all severities, as illustrated by \textit{C-F\textsubscript{VII}} (Avg.\ 0.65, Max.\ 2.56), whose error rises sharply only at the highest severity.
Avg.\ MSE does not capture the full offline effect. Nearly all perturbations (41/42), except \textit{C-D\textsubscript{I}}, exceed a Max.\ dev.\ of 1.046 (about one sixth of a steering-wheel turn), and 17/42 exceed 3.14 (half a steering-wheel turn). Thus, even perturbations with small average error can still induce large instantaneous steering commands, as shown by \textit{C-F\textsubscript{IV}} (Max.\ MSE 0.03, Max.\ dev.\ 1.07).

For the LiDAR-based ADS, the trend bars report inverse retention, so increasing bars indicate lower retention at higher severity. Most perturbations follow this monotonic pattern. 

The affine-transformation family (\textit{L-E}) is reported as an aggregated row in \autoref{tab:rq1-combined}, since its three variants show similar retention, ATE, and latency behaviour. 
Only \textit{L-A\textsubscript{II}} and \textit{L-A\textsubscript{III}} deviate, although both remain among the three most harmful perturbations by Avg.\ Ret. \textit{L-B\textsubscript{VI}}, by contrast, follows the monotonic trend and shows the strongest degradation, with Avg.\ Ret.\ 38.01\% and Min.\ Ret.\ 23.40\%.

Across LiDAR perturbations, Avg.\ Ret.\ and Min.\ Ret.\ are generally aligned, indicating similar average and worst-case degradation. However, retention impact is often small: nearly half of the perturbations (14/31) reduce retention by at most 10\%, while only 11/31 reduce minimum retention below 80\%. ATE is less aligned with retention: \textit{L-B\textsubscript{VI}}, with the lowest Min.\ Ret., ranks only 18th in maximum ATE, while \textit{L-F\textsubscript{II}}, the second-highest perturbation in ATE, ranks only 13th in retention. More broadly, six perturbations exceed 0.5\,m in maximum ATE, whereas ten remain below 0.25\,m. Thus, perturbations can affect retention and localization differently. \\

\rqbox{\textbf{RQ\textsubscript{1} (model-level testing):}
Camera perturbations follow different severity-response patterns and can still trigger sharp steering reactions despite limited overall degradation. LiDAR perturbations mainly reduce detection performance as perturbation strength increases, while their effect on localization is less consistent.
}

\subsubsection{Hardware-in-the-loop feasibility (RQ\textsubscript{2})}

The next two columns of \autoref{tab:rq1-combined} report average and maximum latency for camera and LiDAR perturbations.
For camera-based perturbations, most satisfy the 33\,ms real-time constraint: only 5/42 exceed it, and only 2/42 also exceed 200\,ms, indicating clear infeasibility. Although two infeasible cases rank relatively high in effectiveness (5th and 7th), most of the strongest perturbations remain within real-time limits. Node overhead is negligible for all feasible perturbations, reaching at most 0.71\,ms (\textit{C-C\textsubscript{V}}) and averaging below 0.36\,ms.

For LiDAR perturbations, feasibility is more restrictive. While many remain within the 200\,ms budget, 6/31 exceed it in average latency and 11/31 in maximum latency. Two of the three most effective perturbations are infeasible, but most infeasible cases are not top ranked, indicating that high latency is not generally associated with high impact. The affine-transformation family (\textit{L-E}), reported as an aggregate in \autoref{tab:rq1-combined}, exceeds the real-time limit overall and is not among the strongest LiDAR perturbations.

This contrast is clearest in the weather-related \textit{L-A} family. For snow, \textit{L-A\textsubscript{II}} (3D-Corruptions-AD) and \textit{L-A\textsubscript{III}} (MultiCorrupt) are infeasible, with average latency reaching $\sim$35\,s, whereas \textit{L-A\textsubscript{I}} and our lightweight approximation \textit{L-A\textsubscript{IV}} remain within real-time constraints, ranking 22nd and 12th in effectiveness. A similar trend appears for rain and fog: the heavier variants (\textit{L-A\textsubscript{V}}, \textit{L-A\textsubscript{VII}}, \textit{L-A\textsubscript{VIII}}) are infeasible, whereas the lightweight approximations (\textit{L-A\textsubscript{VI}}, \textit{L-A\textsubscript{IX}}) remain within budget and are also more effective than their heavier counterparts. Node overhead for feasible LiDAR perturbations remains small, reaching at most 2.48\,ms (\textit{L-F\textsubscript{IV}}) and averaging below 2\,ms. \\

\rqbox{\textbf{RQ\textsubscript{2} (hardware-in-the-loop feasibility):}
Most camera perturbations satisfy real-time constraints. For LiDAR, feasibility depends strongly on implementation: several existing weather perturbations are too slow for online use, while the lightweight variants remain within budget and, in some cases, are also more effective.
}

\subsubsection{System-level testing (RQ\textsubscript{3})}

In the last five columns of \autoref{tab:rq1-combined}, we report completion rate (bars), failure type, and CTE for camera and LiDAR perturbations, together with steering jitter (camera) and detection coverage within 5,m (LiDAR) for non-failing runs.

For camera perturbations, the most severe cases produce different failure profiles. \textit{C-C\textsubscript{XII}} and \textit{C-D\textsubscript{III}} mostly fail with \textit{erratic start} (E), only succeeding once at intensities 1 and 2, respectively. \textit{C-F\textsubscript{I}} instead fails with \textit{collision} (C) after partial execution (13--57\% completion), \textit{C-E\textsubscript{V}} transitions from \textit{erratic start} to \textit{understeering} as intensity decreases, and \textit{C-C\textsubscript{XI}} consistently fails with \textit{understeering} after roughly half of the route. Severe perturbations can thus either prevent progress immediately or induce failure later through accumulating trajectory deviation.
Failures are not limited to the strongest perturbations. \textit{C-A\textsubscript{IV}}, \textit{C-E\textsubscript{VI}}, and \textit{C-B\textsubscript{II}} all fail with \textit{oversteering} after similar completion rates, \textit{C-D\textsubscript{II}} fails with \textit{oversteering} after 72\% completion, and \textit{C-B\textsubscript{VII}} fails with \textit{collision} after 96\% completion despite low model-level error. Conversely, several perturbations with substantial model-level impact, including \textit{C-C\textsubscript{VIII}}, \textit{C-E\textsubscript{IV}}, and \textit{C-F\textsubscript{III}}, complete successfully at the highest intensity. \textit{C-C\textsubscript{X}} further illustrates this mismatch, failing only at intensity 5 although its strongest model-level effect occurs at intensity 4.

Even when runs do not fail, CTE and jitter reveal degradation not visible from completion alone. \textit{C-B\textsubscript{II}} shows the highest Avg.\ CTE (1.29) despite low model-level error, and small Avg.\ MSE can still coincide with noticeable trajectory deviation, as seen for \textit{C-F\textsubscript{IV}} and \textit{C-D\textsubscript{I}}. The two metrics also distinguish degradation patterns: high CTE with low jitter indicates smooth but biased trajectories (e.g., \textit{C-B\textsubscript{II}}), whereas moderate CTE with high jitter indicates unstable steering despite successful completion (e.g., \textit{C-F\textsubscript{III}}, jitter 13.13). Comparing successful and failing runs further shows that some failures develop progressively, with higher pre-failure deviation (e.g., \textit{C-E\textsubscript{V}}: 0.75 vs 0.91), whereas others are more abrupt (\textit{C-B\textsubscript{VII}}: 0.83 vs 0.80). Overall, for camera perturbations, model-level effects provide only limited guidance for closed-loop behaviour.

For LiDAR perturbations, failures are more concentrated, with \textit{collision} (C) as the dominant outcome. Among the highest-impact perturbations, \textit{L-B\textsubscript{VI}}, \textit{L-B\textsubscript{III}}, and \textit{L-A\textsubscript{IV}} fail at intensities 2 and 1 after partial completion, while \textit{L-A\textsubscript{IX}} fails at all tested intensities. Unlike the camera model, LiDAR perturbations can therfore still induce failure at the lowest tested intensity.
However, strong detector-level degradation is neither necessary nor sufficient for system-level failure. Some perturbations with comparatively strong model-level effects, such as \textit{L-B\textsubscript{II}}, \textit{L-B\textsubscript{IV}}, and \textit{L-D\textsubscript{II}}, succeed at all intensities. Conversely, perturbations with modest model-level effects can still fail early: \textit{L-C\textsubscript{II}} fails at intensity 1 after only 10\% completion, \textit{L-C\textsubscript{I}} fails at intensities 2 and 1 after 13\% and 24\% completion, and \textit{L-C\textsubscript{III}} fails despite near-nominal model-level predictions. Similar model-level trends can also lead to different outcomes: \textit{L-B\textsubscript{II}} and \textit{L-B\textsubscript{III}} show comparable trend shapes, yet only the latter fails, while \textit{L-D\textsubscript{II}} succeeds at all intensities, whereas \textit{L-D\textsubscript{III}} and \textit{L-D\textsubscript{IV}} fail throughout.

Non-collision failures are less common but still recurring. \textit{L-A\textsubscript{VI}} fails with \textit{understeering} and then \textit{collision}, \textit{L-A\textsubscript{IV}} and \textit{L-B\textsubscript{V}} show mixed \textit{understeering}/\textit{collision}, and \textit{L-C\textsubscript{I}} is the only \textit{road departure} case. Unlike in camera, \textit{erratic start} does not occur: LiDAR runs begin successfully and deteriorate during execution.
Coverage within 5\,m remains high overall, with the lowest value at 0.94, indicating that vehicles are detected in most frames, but this metric does not capture temporal instability. During real runs, vehicles briefly disappeared and reappeared across consecutive frames, suggesting unstable inputs to the planner. This interpretation is supported by higher pre-failure CTE in several cases, most strongly for \textit{L-C\textsubscript{II}} (2.77 to 4.85), and also for \textit{L-A\textsubscript{X}}, \textit{L-C\textsubscript{I}}, \textit{L-B\textsubscript{III}}, \textit{L-B\textsubscript{V}}, \textit{L-B\textsubscript{VI}}, and \textit{L-B\textsubscript{I}}. Under this view, even brief missed detections can be sufficient to cause \textit{collision}, while repeated disappearance and reappearance of obstacles can also contribute to \textit{understeering} and \textit{road departure} by preventing the planner from converging to a stable trajectory. \\

\rqbox{\textbf{RQ\textsubscript{3} (system-level testing):}
Camera and LiDAR perturbations both induce harmful closed-loop effects, but in different ways. For camera, failures vary in type and onset, and even successful runs can show substantial trajectory deviation or steering instability. For LiDAR, failures are more concentrated, predominantly as collisions, and can arise even when detector-level degradation appears small. Overall, model-level effects provide only limited guidance for on-vehicle behaviour.
}

\subsubsection{Perturbation-based fine-tuning (RQ\textsubscript{4})}

\autoref{tab:rq4-relative-to-nominal} reports completion, failure type, and Avg.\ CTE and jitter for both the fine-tuned (FT) and baseline (Non-FT) models across different environmental conditions. For successful runs, these metrics are computed over the full trajectory; for failing runs, they are computed only over the completed portion before failure.

Fine-tuning consistently improves robustness. The FT model completes all runs across all tested conditions, whereas the Non-FT model frequently fails, including \textit{collision} (C), \textit{oversteering} (O), and \textit{road departure} (R).

Under nominal conditions, both models achieve full completion. However, the FT model shows lower Avg.\ CTE (1.02 vs.\ 1.18), lower Avg.\ jitter (0.44 vs.\ 0.64), lower Max.\ jitter (24.91 vs.\ 26.75), and lower jitter variability (1.25 vs.\ 1.47), indicating smoother and more accurate trajectory tracking even without distribution shift.

Differences become more pronounced under adverse conditions. In Wetground, the Non-FT model achieves one full completion, but the other two runs terminate at 42\% and 62\% with \textit{road departure} and \textit{collision}, alongside a large increase in Max.\ CTE (3.09); the FT model completes all runs. In Mud and Raindrops, the Non-FT model never completes the route, with mixed \textit{collision}/\textit{oversteering} failures and completion rates between 43\% and 60\% in Mud and between 35\% and 61\% in Raindrops. In Sun, all Non-FT runs end in \textit{collision}, with one run reaching 67\% completion and the other two failing very early (14\% and 13\%). In all three conditions, the FT model again completes every run.

For perturbed conditions, comparison is restricted to the shared non-failing portion of each run. Even under this conservative comparison, the FT model maintains lower Avg.\ CTE in every condition and lower Avg.\ jitter in all cases except Mud. Although it often shows higher maximum jitter, its lower standard deviation in most perturbed conditions suggests that these peaks are isolated transient corrections rather than sustained unstable behavior. \\

\rqbox{\textbf{RQ\textsubscript{4} (perturbation-based fine-tuning):}
Perturbation-based fine-tuning improves both robustness and stability. It removes failures across all tested conditions and reduces trajectory deviation and steering variability, indicating improved closed-loop behaviour.
}

\begin{table}[t]
\centering
\setlength{\tabcolsep}{3pt}
\renewcommand{\arraystretch}{0.95}
\caption{RQ\textsubscript{4}: Fine-tuning results under nominal and real weather conditions.}
\label{tab:rq4-relative-to-nominal}
\resizebox{\columnwidth}{!}{%
\begin{tabular}{lllccccccc}
\toprule
Weather & SUT & Completion & Fail. Types & \multicolumn{2}{c}{CTE} & \multicolumn{3}{c}{Jitter} \\
\cmidrule(lr){5-6} \cmidrule(lr){7-9}
 & & & & Avg & Max & Avg & Max & Std \\
\midrule
Nominal & FT & \begin{tikzpicture}[baseline={(current bounding box.south) - 0.6ex}]\draw[fill=green!70!black, draw=white, line width=0.1mm] (0*\BARWIDTHfour, 0) rectangle (1*\BARWIDTHfour, 1.00*\BARHEIGHT);\draw[fill=green!70!black, draw=white, line width=0.1mm] (1*\BARWIDTHfour, 0) rectangle (2*\BARWIDTHfour, 1.00*\BARHEIGHT);\draw[fill=green!70!black, draw=white, line width=0.1mm] (2*\BARWIDTHfour, 0) rectangle (3*\BARWIDTHfour, 1.00*\BARHEIGHT);\draw[black] (0,0) -- (3*\BARWIDTHfour,0);\draw[black] (0,1.3*\BARHEIGHT) -- (3*\BARWIDTHfour,1.3*\BARHEIGHT);\end{tikzpicture} & [$\checkmark$,$\checkmark$,$\checkmark$] & 1.02 & 1.11 & 0.44 & 24.91 & 1.25 \\
 & Non-FT & \begin{tikzpicture}[baseline={(current bounding box.south) - 0.6ex}]\draw[fill=green!70!black, draw=white, line width=0.1mm] (0*\BARWIDTHfour, 0) rectangle (1*\BARWIDTHfour, 1.00*\BARHEIGHT);\draw[fill=green!70!black, draw=white, line width=0.1mm] (1*\BARWIDTHfour, 0) rectangle (2*\BARWIDTHfour, 1.00*\BARHEIGHT);\draw[fill=green!70!black, draw=white, line width=0.1mm] (2*\BARWIDTHfour, 0) rectangle (3*\BARWIDTHfour, 1.00*\BARHEIGHT);\draw[black] (0,0) -- (3*\BARWIDTHfour,0);\draw[black] (0,1.3*\BARHEIGHT) -- (3*\BARWIDTHfour,1.3*\BARHEIGHT);\end{tikzpicture} & [$\checkmark$,$\checkmark$,$\checkmark$] & 1.18 & 1.49 & 0.64 & 26.75 & 1.47 \\
\midrule
Wetground & FT & \begin{tikzpicture}[baseline={(current bounding box.south) - 0.6ex}]\draw[fill=green!70!black, draw=white, line width=0.1mm] (0*\BARWIDTHfour, 0) rectangle (1*\BARWIDTHfour, 1.00*\BARHEIGHT);\draw[fill=green!70!black, draw=white, line width=0.1mm] (1*\BARWIDTHfour, 0) rectangle (2*\BARWIDTHfour, 1.00*\BARHEIGHT);\draw[fill=green!70!black, draw=white, line width=0.1mm] (2*\BARWIDTHfour, 0) rectangle (3*\BARWIDTHfour, 1.00*\BARHEIGHT);\draw[black] (0,0) -- (3*\BARWIDTHfour,0);\draw[black] (0,1.3*\BARHEIGHT) -- (3*\BARWIDTHfour,1.3*\BARHEIGHT);\end{tikzpicture} & [$\checkmark$,$\checkmark$,$\checkmark$] & 1.21 & 1.29 & 0.29 & 12.26 & 0.83 \\
 & Non-FT & \begin{tikzpicture}[baseline={(current bounding box.south) - 0.6ex}]\draw[fill=red!80!black, draw=white, line width=0.1mm] (0*\BARWIDTHfour, 0) rectangle (1*\BARWIDTHfour, 0.42*\BARHEIGHT);\draw[fill=red!80!black, draw=white, line width=0.1mm] (1*\BARWIDTHfour, 0) rectangle (2*\BARWIDTHfour, 0.62*\BARHEIGHT);\draw[fill=green!70!black, draw=white, line width=0.1mm] (2*\BARWIDTHfour, 0) rectangle (3*\BARWIDTHfour, 1.00*\BARHEIGHT);\draw[black] (0,0) -- (3*\BARWIDTHfour,0);\draw[black] (0,1.3*\BARHEIGHT) -- (3*\BARWIDTHfour,1.3*\BARHEIGHT);\end{tikzpicture} & [R,C,$\checkmark$] & 1.84 & 3.09 & 0.45 & 13.94 & 1.17 \\
Mud & FT & \begin{tikzpicture}[baseline={(current bounding box.south) - 0.6ex}]\draw[fill=green!70!black, draw=white, line width=0.1mm] (0*\BARWIDTHfour, 0) rectangle (1*\BARWIDTHfour, 1.00*\BARHEIGHT);\draw[fill=green!70!black, draw=white, line width=0.1mm] (1*\BARWIDTHfour, 0) rectangle (2*\BARWIDTHfour, 1.00*\BARHEIGHT);\draw[fill=green!70!black, draw=white, line width=0.1mm] (2*\BARWIDTHfour, 0) rectangle (3*\BARWIDTHfour, 1.00*\BARHEIGHT);\draw[black] (0,0) -- (3*\BARWIDTHfour,0);\draw[black] (0,1.3*\BARHEIGHT) -- (3*\BARWIDTHfour,1.3*\BARHEIGHT);\end{tikzpicture} & [$\checkmark$,$\checkmark$,$\checkmark$] & 1.34 & 1.49 & 0.38 & 19.40 & 1.03 \\
 & Non-FT & \begin{tikzpicture}[baseline={(current bounding box.south) - 0.6ex}]\draw[fill=red!80!black, draw=white, line width=0.1mm] (0*\BARWIDTHfour, 0) rectangle (1*\BARWIDTHfour, 0.43*\BARHEIGHT);\draw[fill=red!80!black, draw=white, line width=0.1mm] (1*\BARWIDTHfour, 0) rectangle (2*\BARWIDTHfour, 0.59*\BARHEIGHT);\draw[fill=red!80!black, draw=white, line width=0.1mm] (2*\BARWIDTHfour, 0) rectangle (3*\BARWIDTHfour, 0.60*\BARHEIGHT);\draw[black] (0,0) -- (3*\BARWIDTHfour,0);\draw[black] (0,1.3*\BARHEIGHT) -- (3*\BARWIDTHfour,1.3*\BARHEIGHT);\end{tikzpicture} & [O,O,C] & 1.78 & 2.26 & 0.27 & 11.35 & 0.76 \\
Sun & FT & \begin{tikzpicture}[baseline={(current bounding box.south) - 0.6ex}]\draw[fill=green!70!black, draw=white, line width=0.1mm] (0*\BARWIDTHfour, 0) rectangle (1*\BARWIDTHfour, 1.00*\BARHEIGHT);\draw[fill=green!70!black, draw=white, line width=0.1mm] (1*\BARWIDTHfour, 0) rectangle (2*\BARWIDTHfour, 1.00*\BARHEIGHT);\draw[fill=green!70!black, draw=white, line width=0.1mm] (2*\BARWIDTHfour, 0) rectangle (3*\BARWIDTHfour, 1.00*\BARHEIGHT);\draw[black] (0,0) -- (3*\BARWIDTHfour,0);\draw[black] (0,1.3*\BARHEIGHT) -- (3*\BARWIDTHfour,1.3*\BARHEIGHT);\end{tikzpicture} & [$\checkmark$,$\checkmark$,$\checkmark$] & 1.12 & 1.53 & 0.36 & 21.06 & 1.07 \\
 & Non-FT & \begin{tikzpicture}[baseline={(current bounding box.south) - 0.6ex}]\draw[fill=red!80!black, draw=white, line width=0.1mm] (0*\BARWIDTHfour, 0) rectangle (1*\BARWIDTHfour, 0.67*\BARHEIGHT);\draw[fill=red!80!black, draw=white, line width=0.1mm] (1*\BARWIDTHfour, 0) rectangle (2*\BARWIDTHfour, 0.14*\BARHEIGHT);\draw[fill=red!80!black, draw=white, line width=0.1mm] (2*\BARWIDTHfour, 0) rectangle (3*\BARWIDTHfour, 0.13*\BARHEIGHT);\draw[black] (0,0) -- (3*\BARWIDTHfour,0);\draw[black] (0,1.3*\BARHEIGHT) -- (3*\BARWIDTHfour,1.3*\BARHEIGHT);\end{tikzpicture} & [C,C,C] & 2.08 & 2.58 & 0.51 & 18.55 & 1.29 \\
Raindrops & FT & \begin{tikzpicture}[baseline={(current bounding box.south) - 0.6ex}]\draw[fill=green!70!black, draw=white, line width=0.1mm] (0*\BARWIDTHfour, 0) rectangle (1*\BARWIDTHfour, 1.00*\BARHEIGHT);\draw[fill=green!70!black, draw=white, line width=0.1mm] (1*\BARWIDTHfour, 0) rectangle (2*\BARWIDTHfour, 1.00*\BARHEIGHT);\draw[fill=green!70!black, draw=white, line width=0.1mm] (2*\BARWIDTHfour, 0) rectangle (3*\BARWIDTHfour, 1.00*\BARHEIGHT);\draw[black] (0,0) -- (3*\BARWIDTHfour,0);\draw[black] (0,1.3*\BARHEIGHT) -- (3*\BARWIDTHfour,1.3*\BARHEIGHT);\end{tikzpicture} & [$\checkmark$,$\checkmark$,$\checkmark$] & 1.47 & 1.61 & 0.36 & 19.58 & 1.03 \\
 & Non-FT & \begin{tikzpicture}[baseline={(current bounding box.south) - 0.6ex}]\draw[fill=red!80!black, draw=white, line width=0.1mm] (0*\BARWIDTHfour, 0) rectangle (1*\BARWIDTHfour, 0.61*\BARHEIGHT);\draw[fill=red!80!black, draw=white, line width=0.1mm] (1*\BARWIDTHfour, 0) rectangle (2*\BARWIDTHfour, 0.35*\BARHEIGHT);\draw[fill=red!80!black, draw=white, line width=0.1mm] (2*\BARWIDTHfour, 0) rectangle (3*\BARWIDTHfour, 0.58*\BARHEIGHT);\draw[black] (0,0) -- (3*\BARWIDTHfour,0);\draw[black] (0,1.3*\BARHEIGHT) -- (3*\BARWIDTHfour,1.3*\BARHEIGHT);\end{tikzpicture} & [C,O,O] & 1.74 & 3.25 & 0.53 & 18.51 & 1.30 \\
\bottomrule
\end{tabular}%
}
\end{table}

\subsection{Threats to Validity}

\head{Internal Validity} 
In a closed-loop system, attributing failures to a single component is not always straightforward. We address this by evaluating the same perturbations across model-level, hardware-in-the-loop, and on-vehicle setups, enabling us to observe how effects evolve from prediction changes to full system behavior. We also relate model-level metrics to driving outcomes such as run success, trajectory deviation, and control stability.

\head{External Validity}
A potential threat is that, for the end-to-end model, training and testing take place on the same route, and all experiments use a single vehicle. We mitigate this by separating training and testing in time and season (autumn vs.\ winter), and by evaluating on a public road rather than a closed track. Because parked cars and other scene elements change across days, including vehicles partially occupying the lane, the correct trajectory can differ substantially between training and testing, reducing the risk of overfitting to a fixed scenario. Although all experiments are conducted in a single scenario, the route captures a complex real-world urban scenario with non-trivial driving conditions. To limit confounding variability, RQ\textsubscript{3} evaluates camera perturbations on the same day in rapid succession, with LiDAR perturbations tested on a separate day, while RQ\textsubscript{4} alternates fine-tuned and non-fine-tuned variants. Finally, we study two popular ADS architectures, namely an end-to-end and a modular stack, and complement synthetic perturbations with real-world conditions~\cite{hendrycks2019benchmarkingneuralnetworkrobustness,dong2023benchmarkingrobustness3dobject}.

\head{Reproducibility}
Real-world experiments are hard to reproduce due to hardware and environmental variability. We mitigate this with recorded datasets, ROS bag replay for model-level and HiL experiments, deterministic training and evaluation, and shared perturbation implementations across execution levels.
\section{Qualitative Analysis}\label{sec:qualitative-analysis}

\begin{figure*}[t]
    \centering
    \includegraphics[width=1\linewidth]{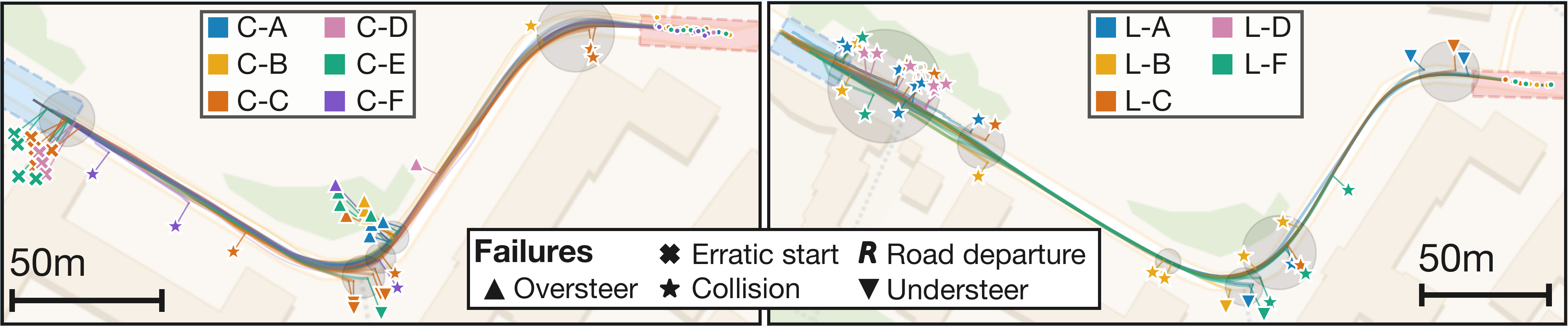}
    \caption{Testing trajectories for camera (left) and LiDAR (right).}
    \label{fig:map_fails}
\end{figure*}

\head{Transfer analysis}
To assess whether model-level effects (RQ\textsubscript{1}) transfer to ViL behavior, we compare offline measurements with system-level outcomes for matched perturbation--intensity pairs that satisfy latency constraints and have valid measurements in both settings. We quantify transfer using (i) Spearman's $\rho$ between offline and ViL measures, with 95\% bootstrap confidence intervals and Benjamini--Hochberg-corrected significance, and (ii) top-$k$ agreement, defined as overlap between the top five perturbation--intensity pairs in the offline and ViL rankings.

For the camera model, transfer is only moderate. Offline MSE is moderately associated with failure rate ($\rho=0.56$, 95\% CI [0.46, 0.64], $q<10^{-4}$) and steering jitter ($\rho=0.50$, 95\% CI [0.25, 0.71], $q<10^{-4}$). Offline maximum steering deviation shows similar associations with failure rate ($\rho=0.55$, 95\% CI [0.44, 0.64], $q<10^{-4}$) and steering jitter ($\rho=0.48$, 95\% CI [0.23, 0.67], $q=0.0001$). However, perturbation-level agreement is weak: top-five overlap is 0.00 for offline MSE versus ViL failure rate, and 0.20 for offline maximum deviation versus ViL steering jitter.

For LiDAR, transfer is weaker and less consistent. Offline retention loss is moderately associated with failure rate ($\rho=0.40$, 95\% CI [0.16, 0.61], $q=0.0041$), but not with 5\,m coverage ($\rho=-0.10$, 95\% CI [-0.42, 0.21], $q=0.5771$). Offline ATE shows only weak association with failure rate ($\rho=0.19$, 95\% CI [-0.06, 0.43], $q=0.2156$) and average cross-track error ($\rho=0.17$, 95\% CI [-0.15, 0.44], $q=0.3865$). Top-five overlap is likewise limited: 0.40 for retention loss versus failure rate, and 0.20 for ATE versus cross-track error.

Overall, offline model-level measurements provide limited guidance for ViL behavior: associations are at most moderate and do not reliably identify the most harmful perturbations in closed-loop execution. Transfer is somewhat stronger for the camera model than for LiDAR, but insufficient in both cases.

\head{Failure hotspots}
We also examine where failures occur during ViL execution. Figure~\ref{fig:map_fails} overlays camera and LiDAR closed-loop trajectories on the route, colors them by perturbation category, and marks failure type. For each failing run, we use the trajectory endpoint as the failure point and cluster these points in the vehicle map frame with DBSCAN, separately by failure type so that nearby but behaviorally distinct events, such as oversteer and understeer, are not merged. This yields failure-type-consistent hotspots; isolated failures remain as noise and are shown in Figure~\ref{fig:map_fails} as semi-transparent circles. For each cluster, we summarize the dominant failure type, dominant and contributing perturbation types.

Figure~\ref{fig:map_fails} shows five recurring camera hotspots, three concentrated in the central route segment. The first, at the route start, is dominated by erratic-start failures, mainly from category~C, with additional contributions from D and E. Mid-route, failures split into adjacent but distinct modes: a large oversteer hotspot involving A, B, C, E, and F; an understeer hotspot driven mainly by C and secondarily by E; and a nearby collision hotspot caused by C and F, reflecting milder understeer cases where the vehicle begins to recover but still drifts into the parked cars after the turn. A final late hotspot near the route endpoint is again collision-dominated, mainly from B and C. Overall, category~C is the most broadly disruptive camera category, spanning start failures, control loss, and collisions, whereas the remaining categories are more localized: B and F mainly contribute to collision-prone regions, E to the central control-loss area, and D almost entirely to the starting hotspot.

LiDAR failures are organized around one dominant early collision hotspot and a smaller set of secondary clusters. The main hotspot appears early in the route, is strongly collision-dominated, and is led by category~A, although B, C, D, and F also contribute, indicating that multiple LiDAR perturbation categories collapse into the same severe early failure mode. A second collision cluster appears later in the route and is mainly associated with category~B, with smaller collision clusters at intermediate completion as well. Unlike the camera case, LiDAR exhibits two understeer hotspots rather than a broad mix of control failures: one near the route end, dominated by category~A with additional contribution from C, and one at mid completion involving A, B, and F. Overall, LiDAR failures are dominated by collision-prone regions, with category~A as the most disruptive category across both early collisions and late understeer, while category~B contributes to later collision hotspots.

\section{Discussion}

\head{Differences across sensors}
Camera perturbations often cause abrupt steering changes, so even small average errors can produce unstable behavior. LiDAR perturbations instead tend to degrade performance more gradually, mainly by reducing detection quality. This difference reflects both sensing modality and system design: the end-to-end model maps inputs directly to control and is therefore sensitive to small changes, whereas the modular stack separates perception and control, leading to more predictable degradation, as also observed in prior work~\cite{survey-lei-ma,sharma2024survey}. This is also reflected in vehicle-in-the-loop failures: camera perturbations produce more diverse control-loss behaviors, whereas LiDAR perturbations are more concentrated in collision-prone regions. For developers, this suggests that robustness evaluation should account for system design: control stability is critical for vision-based systems, while detection quality is central for LiDAR-based systems.

\head{Real-Time constraints matter}
Not all perturbations proposed in the literature, especially for LiDAR, are practically applicable. Some are effective offline but exceed real-time constraints and cannot be used in closed-loop execution. However, many perturbations with strong behavioral impact remain feasible. This shows that real-world robustness testing must consider runtime constraints~\cite{Luan2023Efficient,ayerdi2023metamorphic}.

\head{Effectiveness of perturbation-based training}
Training with perturbed data improves robustness to real-world disturbances such as glare, raindrops, and occlusions in our evaluation. These gains are visible not only in offline metrics but also in closed-loop driving. Importantly, they are observed even when the perturbations are simple and do not closely match real-world effects. Consistent with prior work in data augmentation, this suggests that exposure to diverse perturbations can improve generalization without requiring physically accurate modeling~\cite{cubuk2019randaugment,hendrycks2020augmix}. Hence, even simple perturbations can yield meaningful robustness gains.

\head{Simulation vs. real-world testing}
Offline datasets and simulation are widely used for testing, but both have limitations. Dataset-based evaluation is open-loop and therefore does not capture feedback effects~\cite{hendrycks2019benchmarking,dong2023benchmarkingrobustness3dobject}. Simulation enables closed-loop testing, but relies on simplified sensing and system models, leading to a sim-to-real gap~\cite{deeptest,deeproad,DeepManeuver}. Our transfer analysis shows that offline measurements are at best moderately associated with vehicle-in-the-loop outcomes and do not reliably identify the most harmful perturbations. Our hotspot analysis further shows that failures cluster in specific route segments and modes, which become visible only in real-world closed-loop execution, where perception errors accumulate and interact with control over time. Overall, dataset- and simulation-based testing are necessary but not sufficient; real-world closed-loop evaluation is required to capture system-level effects that do not appear in simplified settings.
\section{Related work}\label{sec:related-work}

\head{Model-level robustness benchmarking}
Prior work has further analyzed the effect of corruption types on generalization, including noise, blur, and environmental degradations~\cite{dodge2016understanding,geirhos2020generalisation,michaelis2020benchmarking}.
In autonomous driving, similar evaluations are conducted for both camera and LiDAR perception. Vision-based robustness is commonly studied using perturbation-based evaluation~\cite{survey-lei-ma,sharma2024survey}, reflecting the broader need for systematic testing methodologies for learning-enabled systems~\cite{RiccioEMSE20,humbatova2020taxonomy}. LiDAR robustness is addressed through benchmarks such as 3D-Corruptions-AD~\cite{dong2023benchmarking, dong2023benchmarkingrobustness3dobject} and MultiCorrupt~\cite{10588664}, which evaluate noise, weather, and geometric distortions in point clouds.
These approaches operate offline and evaluate inputs without capturing temporal effects in closed-loop driving.

\head{System-level and simulation-based testing}
Several works extend perturbation-based testing to system-level evaluation in simulation. DeepXplore~\cite{deepxplore} and DeepTest~\cite{deeptest} apply input transformations to identify erroneous behaviours, while DeepRoad~\cite{deeproad} generates transformed driving scenes using GAN-based methods.
PerturbationDrive~\cite{2025-Lambertenghi-ICST} introduces a comprehensive library of camera perturbations (e.g., noise, blur, weather, geometric distortions, color transformations) and evaluates them in simulation-based closed-loop driving, also exploring perturbation-based retraining. Recent work has further investigated how improvements in simulator realism influence the effectiveness of perturbation-based testing and the resulting system behavior~\cite{2024-Lambertenghi-ICST}.

Other approaches focus on online perturbations or runtime analysis. Adversarial methods~\cite{dataAugment2020Liu,yoon2023learning} generate perturbations during execution to induce failures, while DeepManeuver~\cite{DeepManeuver} applies state-aware perturbations in closed-loop simulation. Runtime monitoring approaches such as Luan et al.~\cite{Luan2023Efficient} and MarMot~\cite{ayerdi2023metamorphic} assess prediction consistency by comparing outputs under controlled input variations.
These works capture temporal effects but are typically limited to simulation and often focus on a single sensing modality.

\head{Real-world robustness evaluation}
Real-world failures of autonomous driving systems highlight the importance of robustness under operational conditions~\cite{damon2018uber,neal2017tesla,david2018uber,stempel2025tesla,jamali2025tesla,reuters2026waymo}. Environmental factors such as adverse weather and sensor noise can significantly degrade perception performance in practice~\cite{hendrycks2019benchmarkingneuralnetworkrobustness,dong2023benchmarkingrobustness3dobject}.
Existing perturbation-based studies are primarily conducted on offline datasets or simulation~\cite{2025-Lambertenghi-ICST,deepxplore,deeptest,deeproad,DeepManeuver,dataAugment2020Liu,yoon2023learning,Luan2023Efficient,ayerdi2023metamorphic}, while real-world analyses rely on observed failures rather than controlled perturbations~\cite{damon2018uber,neal2017tesla,david2018uber,stempel2025tesla,jamali2025tesla,reuters2026waymo}. Recent work has also emphasized the importance of improving the trustworthiness of simulation-based testing before deployment~\cite{biagiola2023better,sorokin2025simulatorensemblestrustworthyautonomous,2025-Baresi-ICSE}.

There is limited work evaluating perturbations in real-world closed-loop driving, particularly across multiple sensing modalities~\cite{2025-lambertenghi-ASE}.
In this work, we evaluate perturbations across both camera and LiDAR modalities using a unified pipeline spanning dataset-level analysis, hardware-in-the-loop testing, and real-world closed-loop driving, enabling consistent evaluation across different levels of system integration.
\section{Conclusions and Future Work}\label{sec:conclusions}

This paper systematically evaluates perturbation-based robustness testing for autonomous driving systems using 72 camera and LiDAR perturbations at the model level, in hardware-in-the-loop, and in vehicle-in-the-loop experiments on a full-scale autonomous vehicle. Across over one million ADS predictions and more than 48 hours of driving on public roads in real-world urban conditions, we study two different ADS stacks: an end-to-end vision-based driving model and a modular LiDAR-based perception, planning, and control stack.

We find that perturbations affect ADS differently across testing levels. Some perturbations with low model-level effects still lead to unstable control, lane departures, or collisions at the system level, while others with strong model-level degradation have only a limited impact in vehicle-in-the-loop driving. This shows that offline robustness does not directly translate to system behavior in practice. These findings underline the need for robustness evaluation across multiple testing levels rather than relying on offline results alone. We further show that real-world perturbation testing is feasible under runtime constraints, and that perturbation-based training improves robustness to naturalistic perturbations while also improving generalization under nominal conditions. Overall, this highlights the practical value of perturbation-based methods for evaluating and improving ADS robustness.

In future work, we will extend the study to more routes, environments, and ADS stacks and explore additional testing and robustness improvement strategies for real-world deployment

\section*{Acknowledgements}
\addcontentsline{toc}{section}{Acknowledgements}
This research was funded by the Bavarian Ministry of Economic Affairs, Regional Development and Energy.

\balance
\bibliographystyle{ACM-Reference-Format}
\bibliography{paper}

\end{document}